\documentclass[12pt,preprint]{aastex}

\newcommand{\myemail}{hayakawa.takehito@jaea.go.jp}
\slugcomment{}
\shorttitle{Empirical Abundance Scaling Laws}
\shortauthors{Hayakawa et al.}

\begin{document}
\title{Empirical Abundance Scaling Laws and Implications for the Gamma-Process in Core-Collapse Supernovae}

\author{TAKEHITO HAYAKAWA\altaffilmark{1,2}, NOBUYUKI IWAMOTO\altaffilmark{3}, 
TOSHITAKA KAJINO\altaffilmark{2,4}, TOSHIYUKI SHIZUMA\altaffilmark{1}, 
HIDEYUKI UMEDA\altaffilmark{4}, KEN'ICHI NOMOTO\altaffilmark{4,5,6}
}

\altaffiltext{1}{Kansai Photon Science Institute, Japan Atomic Energy Agency, Kizu, Kyoto 619-0215, Japan; \myemail}
\altaffiltext{2}{National Astronomical Observatory, Osawa, Mitaka, Tokyo 181-8588, Japan.}
\altaffiltext{3}{Nuclear Data Center, Japan Atomic Energy Agency, Tokai, Ibaraki 319-1195, Japan.}
\altaffiltext{4}{Department of Astronomy, Graduate School of Science, University of Tokyo, Bunkyo-ku, Tokyo 113-0033, Japan.}
\altaffiltext{5}{Department of Astronomy \& Research Center for the Early Universe, School of Science, University of Tokyo, Bunkyo-ku, Tokyo 113-0033, Japan.}
\altaffiltext{6}{Institute for the Physics and Mathematics of the Universe, University of Tokyo, Kashiwa, Chiba 277-8582, Japan.}

\begin{abstract}

Analyzing the solar system abundances,
we have found two empirical abundance scaling laws concerning the $p$- and $s$-nuclei
with the same atomic number.
The first scaling is s/p ratios are almost constant over a wide range of the
atomic number, where the $p$-nculei are lighter than the $s$-nuclei by two or four neutrons.
The second scaling is p/p ratios are almost constant, where 
the second $p$-nuclei are lighter than the first $p$-nucleus by two neutrons.
These scalings are a piece of evidence that most $p$-nuclei are dominantly synthesized by
the $\gamma$-process in supernova explosions.
The scalings lead to a novel concept of "universality of $\gamma$-process"
that the s/p and p/p ratios of nuclei produced by individual $\gamma$-processes
are almost constant, respectively.
We have calculated the ratios by $\gamma$-process based on 
core-collapse supernova explosion models under various astrophysical conditions
and found that the scalings hold for materials produced by individual $\gamma$-processes
independent of the astrophysical conditions assumed.
The universality originates from three mechanisms:
the shifts of the $\gamma$-process layers to keep their peak temperature,
the weak $s$-process in pre-supernovae,
and the independence of the s/p ratios of the nuclear reactions.
The results further suggest an extended universality
that the s/p ratios in the $\gamma$-process layers are not only constant 
but also centered on a specific value of 3.
With this specific value and the first scaling, we estimate that
the ratios of $s$-process abundance contributions from the AGB stars to 
the massive stars are almost 6.7 for the $s$-nuclei of A $>$ 90.
We find that large enhancements of s/p ratios for Ce, Er, and W
are a piece of evidence that the weak $s$-process actually occurred before SNe.

\end{abstract}


\section{INTRODUCTION}

Light elements such as H, He were mainly synthesized in the Big-Bang,
whereas heavier elements were synthesized primarily in many different stellar nucleosynthesis after
the formation of the first generation of stars in the Galaxy.
The solar system was formed from interstellar media (ISM),
which compositions have been provided from the stellar nucleosynthesis.
The solar system abundance
is therefore an important record of these stellar nucleosyntheses and
the galactic chemical evolution (GCE).
The solar system abundance gives several evidence that two neutron capture reaction processes
have happened before the solar system formation.
The first evidence is pairs of two abundance peaks near three-neutron magic numbers of
N = 50, 82, 126.
These two peaks are corresponding to the s- and r-processes, which originated from
the positions of the two nucleosynthesis flows in the nuclear chart \citep{BBFH}.
The second evidence for the s-process is an empirical relation, ${\sigma}{\cdot}N_{s}$ $\sim$ constant,
where ${\sigma}$ and $N_{s}$ are the neutron capture cross-section and the solar abundance 
for the pure $s$-nuclei \citep{Seeger65,kappeler90,Gallino98}.
This can be understood by a concept of "steady flow".
In this way,
the evidence observed in the solar abundances
are correlated with the fundamental mechanisms of the nucleosyntheses.

We here focus attention to origin of $p$-process isotopes ($p$-nuclei).
They have the following features.
First, they cannot be synthesized by the neutron-capture reactions
because they are on the neutron-deficient side from the $\beta$-stability line in the nuclear
chart (see Fig.~\ref{fig:chart}).
Second, their isotopic fractions are small (typically 0.1\%-1\%).
Burbidge {\it et al.} 
suggested that the $p$-nuclei may be synthesized by ($\gamma$,n) or (p,$\gamma$) reactions 
about 50 years ago \citep{BBFH}.
Arnould (1976) proposed the $p$-process in pre-supernova phases,
and Woosley and Howard (1978) proposed the $\gamma$-process in supernovae (SNe).
In these pioneering works, the $p$-nuclei are synthesized mainly by photodisintegration reactions from pre-existing nuclei
affected by the $s$-process in earlier evolutionary states of progenitors.
Woosley and Howard (1978) pointed out that the anti-correlation 
between the photodisintegration reaction rates and the solar abundances for the p-nuclei,
which is a piece of evidence of the $\gamma$-process.
After these works, detailed calculations reproduced the relative solar abundances of most $p$-nuclei
within a factor of 3 \citep{Rayet95, Rauscher02, Arnould03}.

Nevertheless, their origin has long been discussed
with many possible nuclear reactions over the last 50 years, 
and their astrophysical sites have not been identified uniquely.
The proposed nuclear processes are the rapid proton capture reactions in novae or Type I X-ray bursts in
neutron stars (rp-process) \citep{Schatz98,Schatz01},
the proton-induced reactions by Galactic cosmic rays \citep{Audouze70},
the $\gamma$-process in core-collapse SNe \citep{Arnould76,Woosley78,Rayet95}
the explosive nucleosynthesis in Type Ia SNe \citep{Howard91,Arnould03,Kusakabe05},
supernova-driven supercritical accretion disks \citep{Fujimoto03},
and the neutrino-induced reactions in SN explosions 
($\nu$-process) \citep{Woosley90,Hoffman96}.
The origin of the p-nuclei is crucial to our understanding of how the solar-system material formed and evolved.
In our previous paper \citep{Hayakawa04},
we reported two empirical scaling laws found in the solar system abundances,
which are a piece of evidence that the most probable origin of the $p$-nuclei is the SN $\gamma$-process.
The empirical laws lead a novel concept of "the universality of the $\gamma$-process"
that each SN $\gamma$-process arising from different conditions 
in individual SN explosions
should reproduce N(s)/N(p) $\approx$ constant over a wide range of atomic number \citep{Hayakawa04,Hayakawa06a}.
The purpose of this paper is to report detailed analyses of these scalings.
We present the mechanism of this universality
in core-collapse SN $\gamma$-process model calculations.

\section{ANALYSES OF THE SOLAR ABUNDANCES}

\subsection{\it Discovery of the first scaling}

\begin{figure}
\includegraphics[viewport=0mm 0mm 140mm 110mm, clip, scale=0.7]{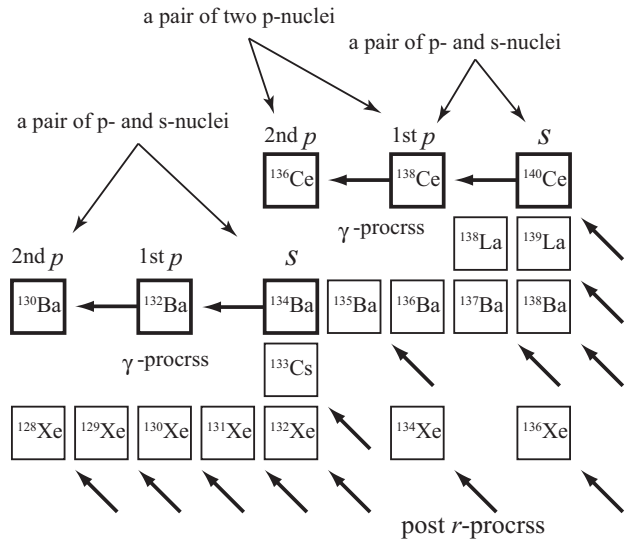}
\caption{The partial nuclear chart around the Xe and Ba isotopes. 
The nuclei located on the $\beta$-stability line are mainly synthesized by the $s$- and $r$-processes.
In contrast, the $p$-nuclei, for example $^{130,132}$Ba and $^{136,138}$Ce,
are synthesized by the $\gamma$-process.
There are pairs of a p-nucleus and an s-nucleus that is  heavier than
the p-nucleus by two neutrons.
}
\label{fig:chart}       
\end{figure}

The $p$-nuclei are on the neutron-deficient site from the $\beta$ stability line
as shown in Fig.~\ref{fig:chart}.
Typical $p$-nuclei are even-even isotopes, of which both proton and neutron numbers
are even.
They are isolated in the nuclear chart, around which isotopes are unstable.
There are twenty-two pairs of a $p$-nucleus and an $s$-nucleus
that is heavier than the $p$-nucleus by two neutrons.
In most cases, the $s$-nuclei are pure $s$-nuclei
that are dominantly synthesized by the $s$-process 
since these $s$-nuclei are shielded by stable isobars against ${\beta}^{-}$-decay
after freezeout of the $r$-process.
Nine elements have two $p$-nuclei.
We here define the first and second $p$-nuclei 
that are lighter than 
the $s$-nucleus by two and four neutrons, respectively.
Figure \ref{fig:chart} shows a partial nuclear chart as a typical example.
$^{134}$Ba is a pure $s$-nucleus shielded by a stable isobar $^{134}$Xe
against the $\beta$-decay after the freezeout of the $r$-process.
The stable isotopes, $^{132}$Ba and $^{130}$Ba, are the first and second $p$-nuclei,
respectively.
It is noted that $^{144}$Sm is the 2nd p-nucleus and
$^{148}$Sm is the $s$-nucleus but the first $p$-nucleus $^{146}$Sm is unstable.
There are thirty-five $p$-nuclei:
twenty-two first $p$-nuclei, ten second $p$-nuclei,
and other three $p$-nuclei.
The other three $p$-nuclei are odd-$N$ isotopes of $^{115}$Sn, $^{138}$La, and $^{180}$Ta
and beyond of the scope of following discussion of the solar abundance ratios.

\begin{figure}
\includegraphics[viewport=0mm 0mm 220mm 170mm, clip, scale=0.45]{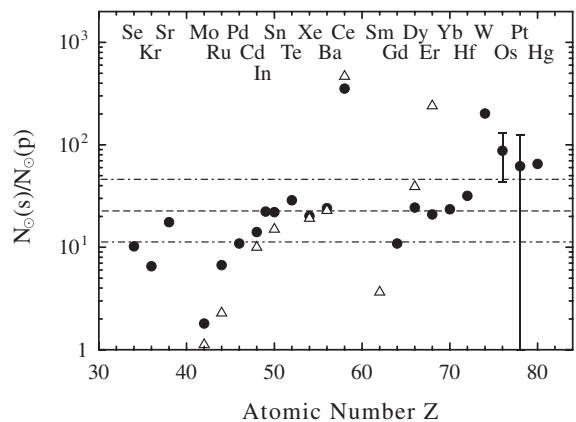}
\caption{Solar abundance ratios of $s$-nucleus to $p$-nucleus with the same atomic number.
The circles and triangles are N$_{\odot}$(s)/N$_{\odot}$(1st p)
and N$_{\odot}$(s)/N$_{\odot}$(2nd p), respectively, where N$_{\odot}$ is the solar abundance.
The first and second $p$-nuclei are lighter than the $s$-nucleus by two and four neutrons, respectively.
The dotted line is the constant value of N$_{\odot}$(s)/N$_{\odot}$(p) = 23. 
The dot-dashed lines are 11.5 and 46.
The observed ratios are almost constant within a factor of 2.
}
\label{fig:sp_solar}
\end{figure}

\begin{figure}
\includegraphics[viewport=0mm 0mm 200mm 160mm, clip, scale=0.45]{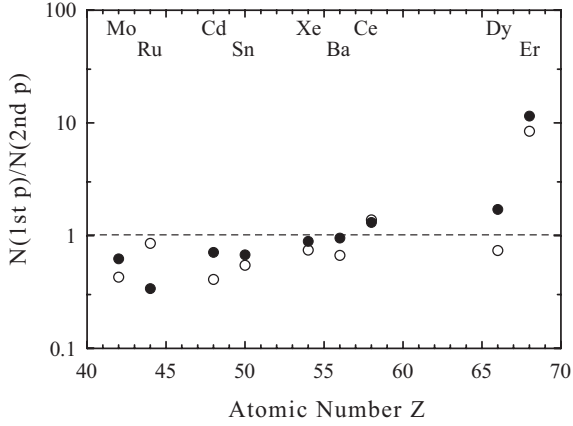}
\caption{Abundance ratios of two pure p-nuclei, N(1st p)/N(2nd p). 
The first and second p-nuclei are respectively
two and four neutron-deficient isotopes from s-nucleus with the same atomic number $Z$.
The filled circles mean the observed ratios in the solar system.
The open circles stand for the calculated ratios
in the model with $M$ = 25 $M_\odot$,   $Z$ = $Z_{\odot}$ and $E$ = 10$^{51}$ erg.}
\label{fig:pp_solar}       
\end{figure}

Here we discuss the isotope abundance ratios
of a $p$-nucleus and an $s$-nucleus with the same atomic number.
Taking the abundance ratios 
of the s-nucleus to the p-nucleus,
N$_{\odot}(s)$/N$_{\odot}$(p), 
where N$_{\odot}$(s) and N$_{\odot}$(p) are the solar isotope abundances of the $s$-
and $p$-nuclei, respectively.
For example, the solar isotope abundances of a $p$-nucleus $^{132}$Ba and 
a $s$-nucleus $^{134}$Ba are 0.101\% and 2.417\%, respectively,
and the isotope abundance ratio of N$_{\odot}$($^{134}$Ba)/N$_{\odot}$($^{132}$Ba) is 23.93.
We reported the first scaling that N$_{\odot}$($s$)/N$_{\odot}$($p$) ratios are almost constant over a wide region
of atomic number \citep{Hayakawa04}.
We also reported another empirical scaling between two p-nuclei 
with the same atomic number as shown in Fig.~\ref{fig:sp_solar}\citep{Hayakawa04}.
Nine elements have two p-nuclei.
As shown in Fig.~\ref{fig:pp_solar},
the N$_{\odot}$(1st p)/N$_{\odot}$(2nd p) ratios are almost constant over a wide region of the atomic number
except a large deviation of Er,
of which the reason will be discussed later.

The N$_{\odot}$(s)/N$_{\odot}$(1st p) and N$_{\odot}$(1st p)/N$_{\odot}$(2nd p) ratios 
show clear correlation \citep{Hayakawa04}, 
but a question whether the ratios of N$_{\odot}$(s)/N$_{\odot}$(2nd p) can be described by the first scaling or not
has been remained.
Here we present both ratios of N$_{\odot}(s)$/N$_{\odot}$(1st p) 
and N$_{\odot}(s)$/N$_{\odot}$(2nd p) in Fig.~\ref{fig:sp_solar}.
There is a clear correlation between them.
The ratios are almost constant over a wide range of the atomic number
except for some deviations, of which the reason will be discussed later.
The ratios in the region of 49 $\le$ Z $\le$ 72 except Ce and Er
is centered around their average of 23.2.
In addition, most ratios over a wide region are constrained at a constant value of
N$_{\odot}$(s)/N$_{\odot}$(p) $\approx$ 23 within a factor of 2.
We summarize the isotope abundances and their ratios in table 1.

The first scaling shows a strong correlation between
p- and s-nuclei with the same atomic number, 
which indicates that the origin of the p-nuclei is strongly correlated with the s-nuclei.
This is consistent with the previous theoretical calculations that 
the p-nuclei are produced by the $\gamma$-process ($p$-process)
in SN explosions \citep{Arnould76,Woosley78,Rayet90,Prantzos90,Rayet95,Rauscher02,Arnould03}.
In the $\gamma$-process models,
pre-existing nuclei in massive stars are affected by the weak s-process
during pre-supernova evolutionary stages.
These pre-existing nuclei originate from early generations of stars
through interstellar media.
The $p$-nuclei are produced from them by photodisintegration
reactions such as ($\gamma$,n) reactions
in a huge photon bath at extremely high temperatures in SN explosions. 
The previous calculations indicated that the $p$-nuclei are
produced via two nuclear reaction paths.
The first is direct ($\gamma$,n) reactions from heavy isotopes.
The second is the EC/${\beta}^{+}$-decay from the neutron-deficient unstable nuclei
after the freezeout of the $\gamma$-process.
These neutron-deficient nuclei are first transmuted
by successive photodisintegration reactions as ($\gamma$,n) reactions
from heavier isotopes
and subsequently transmuted by downflows of ($\gamma$,p) and ($\gamma$,$\alpha$) reactions.
The first scaling suggests that the former reaction path
are likely to play a role more important than the latter reaction path.
The contribution of these two nucleosynthesis paths is quantitatively discussed later.
The charged particle reactions in the rp-process \citep{Schatz98,Schatz01}
and proton-induced reactions by the cosmic rays \citep{Audouze70}
change the proton number of the seed nuclei.
In the $\nu$-process,
the charged current interaction that has a contribution larger than
the neutral current interaction also changes the proton number \citep{Goriely01,Heger05}.
Therefore,
the scaling does not emerge from the dominant charged particle
processes and the $\nu$-process.
The first scaling is, thus, a piece of evidence that the SN $\gamma$-process
is the most probable origin of the $p$-nuclei.

\begin{figure}
\includegraphics[viewport=0mm 0mm 200mm 150mm, clip, scale=0.55]{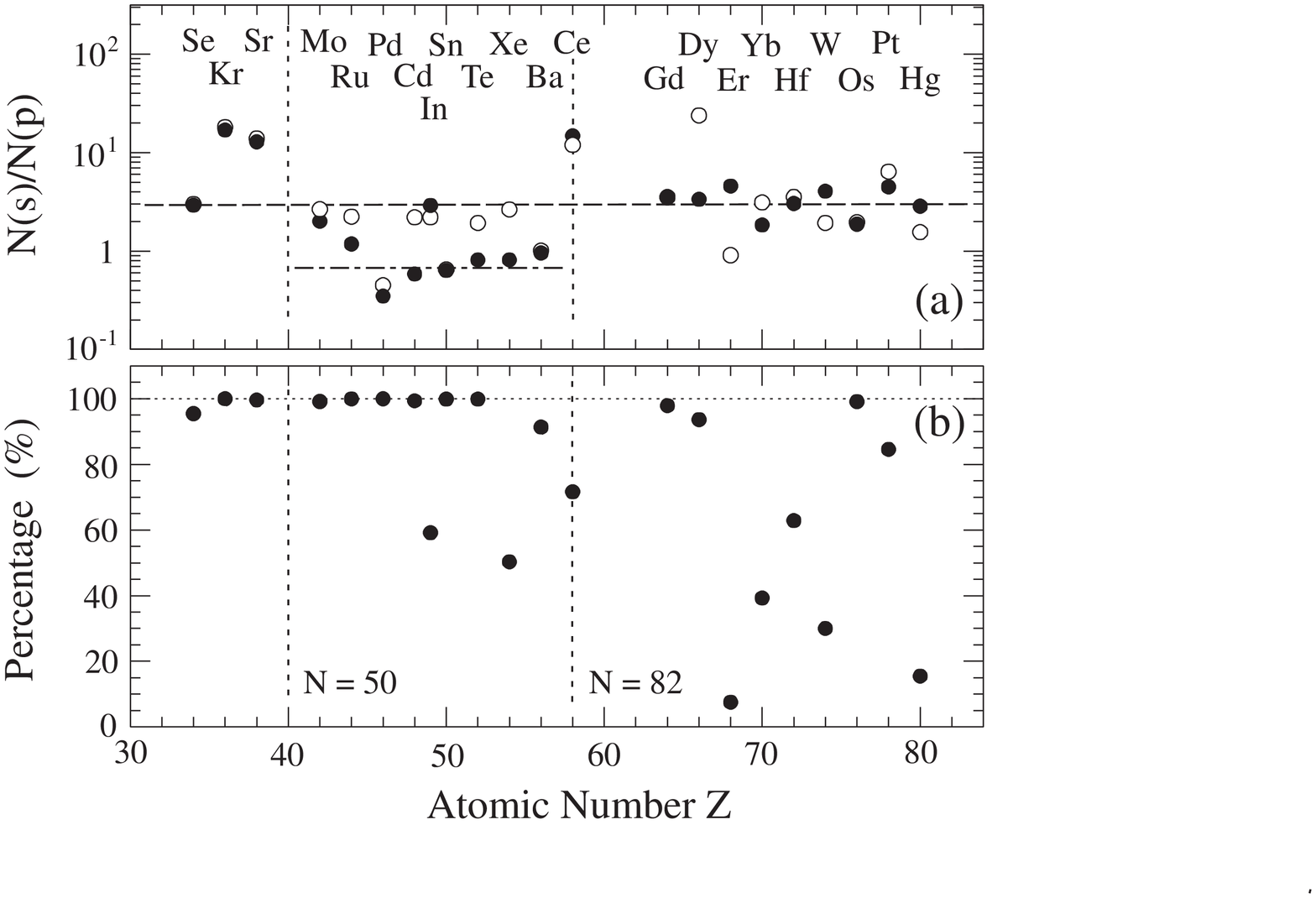}
\caption{Solar abundances of the $p$- and $s$-nuclei, which are the members of the scalings.
The dot-dashed lines are the neutron magic numbers of N=50 and 82. The dot lines mean constant
values around the centers of the $p$- and $s$-nucleus abundances in each region.
}
\label{fig:abundances}       
\end{figure}

Figure \ref{fig:abundances} shows the solar abundances of the $p$-nuclei
and $s$-nuclei that are members of the first scaling.
The correlation between the $p$- and $s$-nuclei
can be observed in this figure.
Seeger et al. (1965) pointed out that the empirical relation,
${\sigma}{\cdot}N_{s}$ $\sim$ constant for the pure $s$-nuclei.
We would like to stress
that the abundances of the $s$-nuclei that are the members of the first scaling, $N_{s}$,
is almost constant in the two mass regions of 50 $<$ $N$ $<$ 82 and 82 $<$ $N$,
respectively.
The abundance pattern of the $p$-nuclei shows a tendency similar to that of the $s$-nuclei.
The abundances of the $p$-nuclei are almost constant in the two mass regions
of 50 $<$ $N$ $<$ 82, and 82 $<$ N, respectively.
There are some enhancements of the $p$-nuclei: for example $^{112}$Sn, $^{196}$Hg.
Their partner $s$-nuclei, $^{116}$Sn and $^{198}$Hg, also show enhancements.
These enhancements show the abundances of the $p$-nuclei are
proportional to those of the $s$-nuclei with the same atomic number.
These are consistent with the first scaling that the N$_{\odot}$(s)/N$_{\odot}$(p) ratios
are almost constant over a wide range of the atomic number.

\subsection{\it Reason for the deviations}

Figure \ref{fig:sp_solar} shows some deviations from the average value of 23.
These deviations can be explained by contributions of other nucleosynthesis
processes or other astrophysical origin.
\\
\\

\noindent
{\bf Ce, Er, and W}:
The three large deviations for Ce, Er and W can be explained by an exceptional contribution
from the r-process because the heavier isotopes in the pairs, $^{140}$Ce, $^{166}$Er, and $^{182}$W, 
are not shielded against ${\beta}^{-}$-decay
after the freezeout of the r-process.
The r-process contributions to $^{166}$Er and $^{182}$W are larger than the s-process contributions
in theoretical calculations \citep{Arlandini99}.
In addition the effect of the neutron magic number N=82 should contribute
to the deviation for Ce.
\\
\\

\noindent
{\bf Gd}: The small deviation for Gd can be explained by a contribution
of a weak branch of the $s$-process.
A unstable nucleus $^{151}$Sm is known as a branching point of the $s$-process,
where the neutron capture reaction and $\beta$-decay compete.
The $p$-nucleus $^{152}$Gd is produced by a following $s$-process branch
\[
\mbox{
$^{150}$Sm(n,$\gamma$)$^{151}$Sm$(e^- \nu)^{151}$Eu(n,$\gamma$)$^{152}$Eu$(e^-\nu)^{152}$Gd.
}
\]
As the result, the N$_{\odot}$($^{154}$Gd)/N$_{\odot}$($^{152}$Gd) ratio 
decreases from the N($^{154}$Gd)/N($^{152}$Gd) ratio at SNe.
\\
\\

\begin{figure}
\includegraphics[viewport=0mm 0mm 160mm 110mm, clip, scale=0.7]{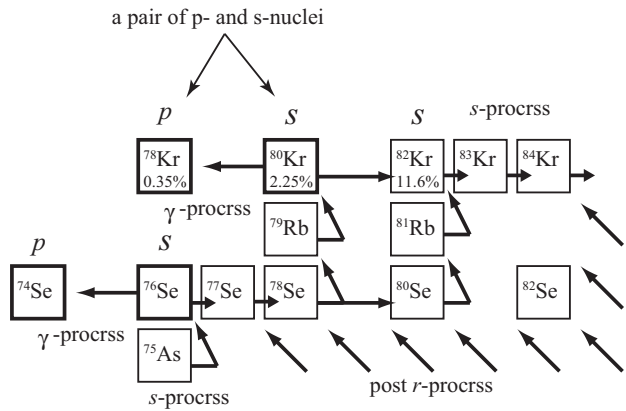}
\caption{A partial nuclear chart and nucleosynthesis flow around Kr isotopes.
$^{80,82}$Kr are pure $s$-nuclei since they are shielded
against the $\beta$-decay after the $r$-process. $^{80}$Kr is synthesized
by a weak branch of the $s$-process, whereas $^{82}$Kr is located on the
main flow of the $s$-process.
}
\label{fig:Kr}       
\end{figure}

\noindent
{\bf Kr}: 
The deviation of N$_{\odot}$($^{80}$Kr)/N$_{\odot}$($^{78}$Kr) may originate from a weak branch of the $s$-process.
As shown in Fig.~\ref{fig:Kr}, $^{79}$Se is a branching point
and $^{80}$Kr is synthesized by a weak branch of the $s$-process.
$^{82}$Kr is located on the main flow of the $s$-process
and the isotope abundance of $^{82}$Kr is about 4 times larger than
that of $^{80}$Kr.
If we take a ratio N$_{\odot}$($^{82}$Kr)/N$_{\odot}$($^{78}$Kr) instead of N$_{\odot}$($^{80}$Kr)/N$_{\odot}$($^{78}$Kr),
it is about 33.
\\
\\

\noindent
{\bf Mo and Ru}:
Four isotopes of $^{92,94}$Mo and $^{96,98}$Ru located 
near the neutron magic number of $N$=50 show large deviations.
Isotope abundances of typical $p$-nuclei in a mass region around Mo and Ru
are 0.3-1\%, whereas the isotope abundances of $^{92,94}$Mo and $^{96,98}$Ru
are 14.84, 9.25, 5.54, 1.87\%, respectively.
Thus, the ratios of Mo and Ru are lower than the average value of 23 by an order of magnitude.
Previous $\gamma$-process (or $p$-process) model calculations 
tried to reproduce the relative abundances of all the $p$-nuclei.
However, the calculated results showed the underproduction
of Mo and Ru. \citep{Rayet95}.
The large deviations in the N$_{\odot}$(s)/N$_{\odot}$(p) ratios 
are consistent with the previous calculated results
and suggest that their origin may be different with the other $p$-nuclei.
\\
\\

\noindent
{\bf Sm}:
The N$_{\odot}$($^{148}$Sm)/N$_{\odot}$($^{144}$Sm) ratio shows large deviation
like those of Mo and Ru. The $p$-nucleus $^{144}$Sm is on the neutron magic number of $N$=82.
This situation is similar to Mo and Ru
and the large deviation suggests that the origin of $^{144}$Sm might be same as
those of Mo and Ru.
\\
\\

\noindent
{\bf Os and Pt}:
The uncertainties for $^{184}$Os and $^{190}$Pt are about 50 and 100\%, respectively\citep{Isotope}.
These deviations may originate from the large uncertainties of the measured abundances.
The measurements of their abundances with a high precision are desired.
\\
\\

\begin{figure}
{\includegraphics[viewport=0mm 0mm 220mm 170mm, clip, scale=0.5]{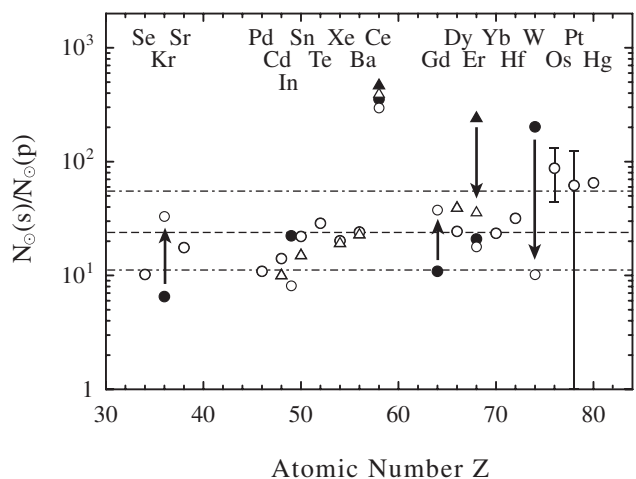}}
\caption{Solar abundance ratios of $s$-nucleus to $p$-nucleus with the same atomic number,
N$_{\odot}$(s)/N$_{\odot}$(p), where N$_{\odot}$ is solar abundance.
The open circles and open triangles are modified N$_{\odot}$(s)/N$_{\odot}$(1st p)
and N$_{\odot}$(s)/N$_{\odot}$(2nd p), respectively.
The filled circles and filled triangles are original N$_{\odot}$(s)/N$_{\odot}$(1st p)
and N$_{\odot}$(s)/N$_{\odot}$(2nd p), respectively.
In most cases, the modified and original ratios are same.
The dotted line is the average of these ratios N$_{\odot}$(s)/N$_{\odot}$(p) = 23. 
The dot-dashed lines are 11.5 and 46.
The observed ratios are almost constant within a factor of 2.
}
\label{fig:sp_solar_mod}
\end{figure}

Figure \ref{fig:sp_solar_mod} shows the corrected N$_{\odot}$(s)/N$_{\odot}$(p) ratios.
The open and filled circles are the original N$_{\odot}$(s)/N$_{\odot}$(p) ratios
and the corrected ratios, respectively.
The $s$-process abundances of $^{140}$Ce, $^{168}$Er, and $^{182}$W are taken from calculated results
in a stellar $s$-process model \citep{Arlandini99}. The $s$-process abundance
for $^{152}$Gd is taken from a calculated result based on neutron capture
reaction rates measured at a n\_TOF facility \citep{nTOF}.
For Kr, we take N$_{\odot}$($^{82}$Kr)/N$_{\odot}$($^{78}$Kr) instead of N$_{\odot}$($^{80}$Kr)/N$_{\odot}$($^{78}$Kr).
To focus our attention to the $\gamma$-process,
we omit the ratios of Mo, Ru, and Sm.
Most corrected ratios are centered around 23 
within a factor of 2 (see Fig.~\ref{fig:sp_solar_mod}).
It is noted that the ratio of In decreases after the correction of the $r$-process contribution.
$^{115}$In does not shielded against the $\beta$-decay
and produced by the $r$-process as Ce, Er, and W
and thus the N($^{115}$In)/N($^{113}$In) ratio decreases. 
This suggests that $^{113}$In 
may have contaminations of other processes as the $s$-process.

\subsection{\it Universality of the $\gamma$-process}

The universality of nucleosynthesis processes is an important concept for understanding
the stellar nucleosyntheses.
The solar system was formed from interstellar media (ISM) originated from 
many stellar nucleosynthesis episodes in the Galaxy.
The stellar nucleosynthesis environments such as mass, metallicity and explosion energy are different
and hence the abundance distributions of the synthesized nuclides may be different.
However, astronomical observations for very metal-deficient stars
have reported the "universal" abundance distributions for Z $>$ 56 \citep{Sneden98,Sneden00},
which are in agreement with the abundance distribution of the r-process nuclides in the solar system.
These facts suggest a uniform site and/or uniform conditions for the synthesis 
of the r-process nuclei.
Otsuki et al. could success to explain the universality of the r-process
by a neutrino-energized wind model \citep{Otsuki00}.

The p- and s-nuclei were produced in different stellar
environments.
Thus, the mass distribution of the synthesized nuclei may depend on the astrophysical
conditions.
Nevertheless, the observed $N_{\odot}$(s)/$N_{\odot}$(p) ratios in the solar system
are almost constant over a wide range of the atomic number.
This leads to a novel concept, "the universality of the $\gamma$-process", 
that the N($s$)/N($p$) ratios of nuclides
produced by individual $\gamma$-processes are constant over a wide region, respectively.
We would like to stress that the concept of the universality of the $\gamma$-process
has been derived from only the analysis of the solar abundances
independent of nucleosynthesis calculations.

The situation of the $\gamma$-process is similar with
that of the s-process.
It is known that 
 ${\sigma}{\cdot}N_{s}$ is roughly constant for the s-process over a large range in nuclear mass,
where ${\sigma}$ and $N_{s}$ are the neutron capture cross-section 
and the solar abundance \citep{Seeger65,kappeler90,Gallino98}.
In this case, one could infer a "universality of the s-process" that arose from the conditions 
being such that a steady neutron capture flow was achieved.  
The difference of the ${\sigma}{\cdot}N_{s}$ for A$<$90 and A$>$90 then pointed to two different s-process components. 
Smaller deviations from constant ${\sigma}{\cdot}N_{s}$ signal interesting nuclear phenomena as weak branch. 
Also, if we could observe abundances of neighboring s-only isotopes in stars of various ages, 
one would expect their abundance ratio to be inversely proportional to the cross section ratio, 
just as they are in the solar system.
The universality of the $\gamma$-process suggests 
that one could find roughly constant N(s)/N(p) ratios
with observation of abundances of p- and s-nuclei in stars.

The universality of the $\gamma$-process suggests
three possible mechanisms
as the reason why the scaling appears in the solar system.
First, a single supernova occurred nearby the solar system before the solar system formation
and strongly affects the solar system materials.
Second, the $\gamma$-process occurs only under a specific uniform astrophysical condition
in the Galaxy 
and thus the mass distributions of materials produced by individual $\gamma$-processes
are almost same.
Third, the $\gamma$-processes under various astrophysical conditions can occur
but N(s)/N(p) ratios are almost constant over a wide region of the atomic number
independent of the astrophysical conditions.
In the last case, a question why 
the scaling holds for materials produced by individual $\gamma$-processes
independent of the astrophysical conditions
has been remained.
The universality is, thus, essential for understanding the nature of the $\gamma$-process
and to constrain the nucleosynthesis site of the $\gamma$-process.

\section{SUPERNOVA MODEL CALCULATIONS AND PRINCIPLE OF THE UNIVERSALITY}
\subsection{\it $\gamma$-process in core-collapse supernova models}

The solar abundances provide two empirical scalings,
which are a piece of evidence that the most probable origin of the $p$-nuclei
is $\gamma$-process in SNe.
A question we have to ask here is whether a standard $\gamma$-process model
in core-collapse SNe
can reproduce the scalings.
In previous studies \citep{Rayet90,Rayet95}, calculated results
were presented by a form of F(x)/F$_0$(x),
where F(x) is the ratio of the calculated abundance to the solar abundance for an isotope x,
and F$_0$ is the average of F(x).
However, these results have not been presented by a form of N(s)/N(p).

We present the calculated N(s)/N(1st p) ratios under various astrophysical conditions.
A massive star evolves from the main sequence to the core-collapse stage and explodes \citep{Nomoto}.
The solar abundances are adopted as the initial composition in the massive star
and is affected by the weak $s$-process in evolutional states.
The mass distribution of the seed nuclei affected by the weak $s$-process
are used as the initial composition at the SN explosions.
Most of the reaction rates are taken from a common astrophysical nuclear data library of REACLIB \citep{REACLIB}.
Figure \ref{fig:sp_cal} shows the results in four different models.
They are a standard model (1) with $M$ = 25 $M_\odot$,   $Z$ = $Z_{\odot}$ and $E$ = 10$^{51}$ erg,
a heavier progenitor mass model (2) with $M$ = 40 $M_\odot$,   $Z$ = $Z_{\odot}$ and $E$ = 10$^{51}$ erg,
a metal-deficient model (3) with $M$ = 25 $M_{\odot}$, $Z$ = 0.05 $Z_{\odot}$ and $E$ = 10$^{51}$ erg,
and a higher energy explosion model (4)
with $M$ = 25 $M_{\odot}$, $Z$ = $Z_{\odot}$ and $E$ = 20 $\times$ 10$^{51}$ erg.
The calculated ratios in the four models
are almost constant over a wide region of $Z$ $>$ 40, respectively.

We have presented that there are three possibilities
as the principle of the universality of the $\gamma$-process.
The calculated results show that even should the $\gamma$-process
occur under the various astrophysical conditions, 
the N(s)/N(p) ratios in individual nucleosynthesis episodes
result in an almost constant value over a wide range
independent of the SN conditions assumed.

\subsection{{\it $\gamma$-process layers correlated with astrophysical conditions}}

Our calculated results have verified the universality of the $\gamma$-process.
However, the reason how the scalings hold for the different nucleosynthesis episodes
has been an open question.
Here we discuss the mechanism of the universality of the $\gamma$-process
in detailed calculations.
The supernova explosion is a dynamical phenomenon and
physical conditions,
for example temperature or density,
 in individual layers of a progenitor are different.
Thus, the physical condition of each $\gamma$-process layer is a key for understanding
the universality of the $\gamma$-process.
It is known that
the mass regions of the $\gamma$-process layers depend on 
the astrophysical conditions and 
are shifted to the outer layers as the explosion energy
or the mass of the progenitor increases \citep{Arnould76,Woosley78,Rayet90,Rayet95}.
In the present calculations, 
the mass regions in the $\gamma$-process models are 1.97 - 2.8 $~M_\odot$ for $10^{51}$ ergs and 2.87 - 4.63~$M_\odot$ 
for 20 $\times$ $10^{51}$ ergs, respectively.
The density and temperature of the layers at the peak of the explosion 
are 
(1.7-3.5) $\times$ 10$^9$ $K$ and (0.13-0.95) $\times$ 10$^6$ g/cm$^3$, respectively, for the model (1),
(1.7-3.5) $\times$ 10$^9$ $K$ and (0.12-0.90) $\times$ 10$^6$ g/cm$^3$ for (2),
(1.6-3.5) $\times$ 10$^9$ $K$ and (0.14-2.1) $\times$ 10$^6$ g/cm$^3$ for (3),
and (1.7-3.5) $\times$ 10$^9$ $K$ and (0.03-0.21) $\times$ 10$^6$ g/cm$^3$ for (4).
Although the mass regions are different,
the temperature ranges are approximately equal to each other.
The shift also is confirmed  in our calculations
and contributes to the universality.

\subsection{{\it Proportionality between the $p$- and $s$-nuclei in the solar system}}

\begin{figure}
{\includegraphics[viewport=0mm 0mm 210mm 260mm, clip, angle=-90,scale=0.3]{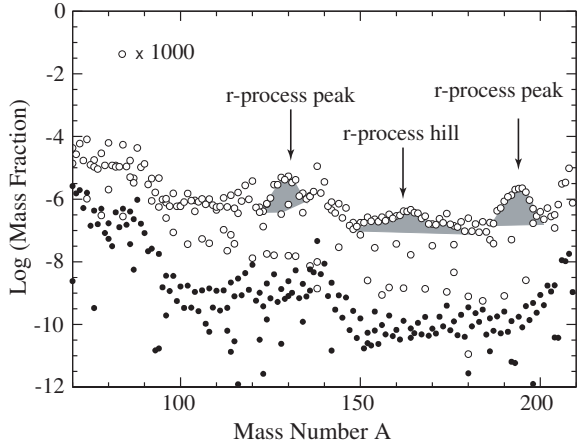}}
\caption{The abundance distribution of the seed nuclei for the $p$-process
after the weak $s$-process (filled circles), and 
as the initial composition (open circles)
in a model with $M$ = 25 $M_\odot$, $Z$ = 0.02.
The open circles have been multiplied by 1000. }
\label{fig:weak_s}
\end{figure}

\begin{figure}
{\includegraphics[viewport=0mm 0mm 220mm 70mm, clip, scale=0.6]{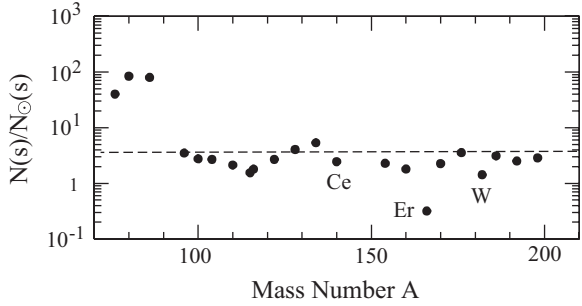}}
\caption{
The abundance ratios of the $s$-processed nuclei
to the initial compositions
for 22 nuclei, which are members of the first scaling.}
\label{fig:ss}
\end{figure}

The $p$-nuclei are dominantly produced by SNe,
whereas the $s$-nuclei are continuously produced by the main $s$-process 
in asymptotic giant branch (AGB) stars \citep{Gallino98,Busso99}.
The question we have to ask here is why the abundances of the $p$-nuclei
is proportional to the $s$-nuclei in the solar system.
A weak $s$-process in evolutional stages of massive stars
gives a hint to answer this question.
It is well known that the weak $s$-process
produces $s$-nuclei of $A$ $<$ 90 from iron seeds.
However, we focus our attention to the weak $s$-process effect to the heavy nuclei of $A$ $>$ 90.
The pre-existing nuclei in the mass region of A $>$ 90 are also
irradiated by successive neutrons in the weak $s$-process
and thereby their abundance pattern is changed.
We calculate 
a mass distribution by the weak $s$-process
in evolutional stages of a massive star
with $M$ = 25 $M_{\odot}$, $Z$ = $Z_{\odot}$.
The open and filled circles in Fig.~\ref{fig:weak_s} show the solar abundances as the initial abundances
adopted in the massive star and the abundances of the s-processed seed nuclei, respectively.
The abundance distribution of the mass region of $A$ $>$ 90 is locally
changed to that of the main $s$-process in the AGB stars.
For example, there are two $r$-process abundance peaks around $A$ = 130 and 195
and a $r$-process hill around $A$=150 (open circles in Fig.~\ref{fig:weak_s})
but these peaks and hill clearly disappear after the weak $s$-process
(filled circles in Fig.~\ref{fig:weak_s}).
To clarify the effect of the neutron irradiation,
we present the abundance ratios of the $s$-processed seeds
to the solar abundances, N(s)/N$_{\odot}$(s),  for the twenty-two $s$-nuclei in Fig.~\ref{fig:ss}.
These $s$-nuclei are the members of the first scaling.
The ratios are
almost constant over the wide region of $A$ $>$ 90; namely, 
the mass distribution of the $s$-processed nuclei of $A$ $>$ 90
is approximately equal to the AGB $s$-process one.
The abundances of the $p$-nuclei are proportional to those of the seed nuclei at the SN explosions.
Therefore the abundance change by the weak $s$-process is the reason for
the proportionality between the $p$- and $s$-nuclei.

The N$_{\odot}$(s)/N$_{\odot}$(p) ratios in the mass region of $A$ $<$ 90 
are lower than the average value of 23 (see Fig.~\ref{fig:sp_solar}).
This is understood by taking into account the enhancement of the s-processed
seed in the mass region of $A$ $<$ 90.
The abundance ratios of N(s)/N$_{\odot}$(s) of $A$ $<$ 90
are larger than those of $A$ $>$ 90 as shown in Fig.~\ref{fig:ss}
and the abundances of the p-nuclei are proportional to those of the massive star $s$-nuclei.
Therefore, the N$_{\odot}$(s)/N$_{\odot}$(p) ratios of $A$ $<$ 90 is lower than those of the higher mass region.

It is noted that there are three deviations for Ce, Er, W
in Fig.~\ref{fig:ss}.
The solar abundances of $^{140}$Ce, $^{166}$Er, and $^{182}$W
are contributed from both the $s$- and $r$-processes since
they are not shielded against ${\beta}^{-}$-decay.
The partial abundances originated from the $r$-process 
disappear after the weak $s$-process
and hence the N(s)/N(p) ratios decrease.

\subsection{{\it Contributions of two nuclear reaction paths}}

Here we study effects of two nucleosynthesis paths in the $\gamma$-process \citep{Rayet90,Rapp06,Rauscher06}.
The first path is direct ($\gamma$,n), ($\gamma$,p), ($\gamma$,$\alpha$) reactions from heavier isotopes.
The ($\gamma$,n) reactions are dominant near the $\beta$ stability line.
The second path is ${\beta}^{+}$ decay after the freezeout of the $\gamma$-process.
The seed nuclei are transmuted to light isotopes by successive photodisintegration reactions.
This reaction flow reaches to neutron-deficient region 
where the ($\gamma$,$\alpha$) or ($\gamma$,p) reaction rate is larger than the ($\gamma$,n) reaction rate. 
The neutron-deficient nuclei 
are subsequently transmuted to lighter elements by downflow of the ($\gamma$,$\alpha$) and ($\gamma$,p) reactions.
After the freezeout of the $\gamma$-process, 
the nuclei on the neutron-deficient side
decay to the stable isotopes.

Woosley and Howard (1978) reported 
the anti-correlation between the photodisintegration reaction rates and the solar abundances 
of the $p$-nuclei, and the ($\gamma$,n) reactions have essential role
for the synthesis of the $p$-nuclei.
The first scaling suggests that the contribution of the first path is dominant.
However, the ${\beta}^+$ decay from the neutron-deficient nuclei may break the scaling
since the mass distribution on the neutron-deficient side flow depends 
on nuclear properties such as particle separation energies.

\begin{figure}
{\includegraphics[viewport=0mm 0mm 220mm 170mm, clip, scale=0.4]{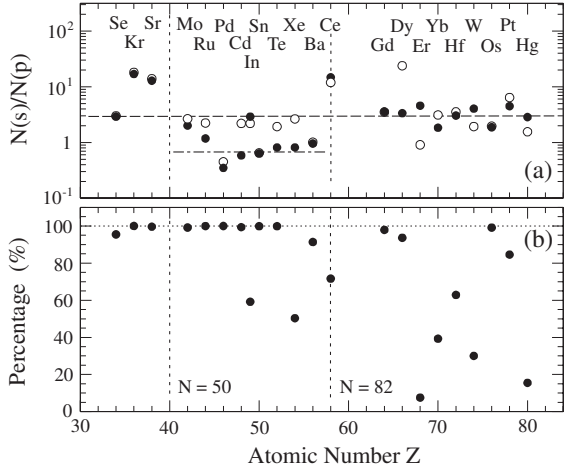}}
\caption{The upper panel presents the calculated N(s)/N(p) ratios.
The filled and open circles are before and after the ${\beta}^{+}$ decay of the neutron-deficient nuclei
in the second path, respectively.
The dashed and dot-dashed line is 3 and 0.7, respectively.
The lower panel presents the percentage of the first paths.
These are calculated in a model with $M$ = 25 $M_\odot$, $Z$ = $Z_{\odot}$ and E = 10$^{51}$ erg.}
\label{fig:paths}
\end{figure}

Here we study an effect of the second path
to the final N(s)/N(p) ratios.
Figure~\ref{fig:paths} (a) shows the N(s)/N(p) ratios before and after the $\beta$-decay of the neutron-deficient
nuclei on the second path. The ratio before the $\beta$-decay is corresponding to that contributed from only the first path.
Figure~\ref{fig:paths} (b) shows percentages of the $p$-nuclei populated from only the first path
to the final abundance due to both the paths.
These N(s)/N(p) ratios and percentages are results
using a model of $Z$ = $Z_{\odot}$, $M$ = 25 $M_\odot$, $E$ = $10^{51}$ erg.

In the mass region of 58 $<$ $Z$ $<$ 82 corresponding to 82 $<$ $N$,
the second path flow is located on the neutron-deficient side
from the $\beta$ stability line by mass unit ${\Delta}A$ $\approx$ 2-6, typically.
Thus, the $p$-nuclei in this mass region are produced by the two reaction paths.
The calculated results show that 
the N(s)/N(p) ratios before the ${\beta}^{+}$ decay 
are almost constant in this mass region (filled circles in Fig.~\ref{fig:paths} (a) ).
However,
the percentages of the first path scatters over a wide range of 5-100\%
(see Fig.~\ref{fig:paths} (b)).
This indicates that the contributions of the second path to the individual $p$-nuclei are
different.
In fact the variance of the final N(s)/N(p) ratios after the $\beta$-decay (open circles in Fig.~\ref{fig:paths} (a) ) 
is larger than that before the ${\beta}^{+}$ decay (filled circles).
However,
the final N(s)/N(p) ratios are still almost constant
because both the p- and s-nuclei are populated by $\beta$-decay.
Therefore the second path does not drastically change the ratios in this mass region.

In the mass region of 40 $<$ $Z$ $<$ 58 corresponding to 50 $<$ N $<$ 82, 
the nucleosynthesis flow mostly proceeds along the $\beta$ stability line.
A similar result was previously reported \citep{Rayet90}.
The ratio of ($\gamma$,n) reaction rates to (n,$\gamma$) reaction rates
are important for the position of the second path in the nuclear chart.
Since the neutron separation energy increases with decreasing the proton member,
the ($\gamma$,n) reaction rates in lower mass regions are lower than those in higher mass regions.
Therefore, the position of the second path in the lower mass regions
is close to the $\beta$-stability line.
The second path proceed mostly on the $\beta$-stability line
in the mass region of $N$ $<$ 82.
The N(s)/N(p) ratios before and after $\beta$-decay 
show almost constant values, respectively.
The N(s)/N(p) ratios before the ${\beta}$-decay (filled circles)
are lower than those after the ${\beta}$-decay (open circles).
This is because the second path locates almost 
on the $\beta$-stability line but some neutron-deficient isotopes
are still produced.
The $\beta$-decay from the neutron-deficient isotopes populates
the $s$-nuclei,
whereas the $p$-nuclei are dominantly produced via the first path (see Fig.~\ref{fig:paths} (b)).
Therefore the ${\beta}$-decay increases the N(s)/N(p) ratios.

In the mass region of 32 $<$ $Z$ $<$ 40,
there are only three elements of Se, Kr and Sr.
The ratios of three elements scatter.
The ratio before the $\beta$-decay are approximately equal to that 
after the $\beta$-decay
since the second path locates almost on the $\beta$-stability line
and the $p$-nuclei are dominantly produced via the first path.

In this section, we have discussed the contributions of the two nucleosynthesis paths.
In most cases in the mass region 40 $<$ $Z$ $<$ 58,
the $\beta$-decay of the second path nuclei populates only the $s$-nuclei in the pairs
and hence 
the N(s)/N(p) ratios increase after the $\beta$-decay.
We would like to stress that
the final ratios in the two mass regions of 40 $<$ $Z$ $<$ 58 and 58 $<$ $Z$ $<$ 82
arr almost constant.
In this way, the second path contributes the manifestation of the first scaling
that the ratios are almost constant over a wide range of the atomic number.
Therefore both the contributions of the first and second paths
have essential role for the first scaling.
It should be noted that
($\gamma$,n) reactions in the two mass region of $N$ $<$ 50 and 50 $<$ $N$ $<$ 82
have dominant role and 
experimental measurements of ($\gamma$,n) reaction rates
on neutron-deficient isotopes (for example, Mohr et al. 2000)
are of importance for the quantitative calculation of 
the $\gamma$-process and the scalings.

\subsection{Effects of ($\gamma$,$\alpha$) reactions}

\begin{figure}
{\includegraphics[viewport=0mm 0mm 200mm 160mm, clip, scale=0.45]{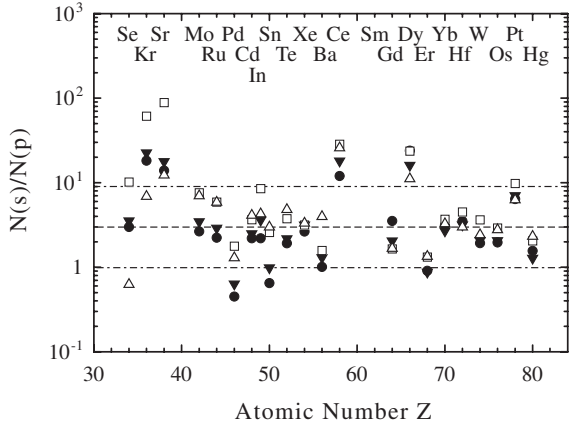}}
\caption{Calculated ratios, N(s)/N(1st p).
The filled circles are the calculated ratios in the model 
with $M$ = 25 $M_\odot$,   $Z$ = $Z_{\odot}$ and $E$ = 10$^{51}$ erg.
The filled triangles are those in the model 
with $M$ = 40 $M_\odot$,   $Z$ = $Z_{\odot}$ and $E$ = 10$^{51}$ erg.
The open triangles are those in the model 
with $M$ = 25 $M_{\odot}$, $Z$ = $Z_{\odot}$ and $E$ = 20 $\times$ 10$^{51}$ erg.
The open squares are those in the model 
with $M$ = 25 $M_{\odot}$, $Z$ = 0.05 $Z_{\odot}$ and $E$ = 10$^{51}$ erg.
The dashed line is N(s)/N(p) = 3.
The dot-dashed lines are 1 and 9.
The ratios in all the models are centered around 3 within a factor of 3 over a wide range
of atomic number for $Z$ $>$ 40.}
\label{fig:sp_cal}
\end{figure}

\begin{figure}
\includegraphics[viewport=0mm 0mm 160mm 130mm, clip, scale=0.5]{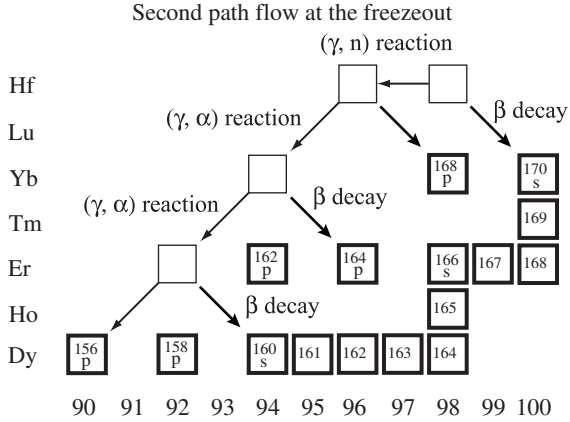}
\caption{A partial nuclear chart around Dy, Er and Yb isotopes and 
a schematic view of nucleosynthesis flows.
The $p$-nuclei are synthesized by two nuclear reactions paths.
The first is a direct photodisintegration reactions such as ($\gamma$,n) reaction from heavier isotopes
and the second is a flow located in neutron deficient side and subsequent $\beta$-decay.
Arrows show a typical second path flow.}
\label{fig:second_paths}
\end{figure}

The calculated N(s)/N(p) ratios in Fig.~\ref{fig:sp_cal} show some deviations. We here discuss the reason of two deviations for Dy and Er 
in viewpoint of the nuclear data input for the nucleosynthesis flow.
Figure \ref{fig:second_paths} shows a typical second path flow at the freezeout of the $\gamma$-process.
In this mass region, the neutron-deficient nuclei are mainly transmuted to lighter elements
by ($\gamma$,$\alpha$) reactions. 
After the freezeout, the nuclei on the second path
decay to the $p$- and/or $s$-nuclei. 
In most cases, both the $p$- and $s$-nuclei have the contributions
from the $\beta$-decay and the final N(s)/N(p) ratios are not drastically changed 
from the ratios before the $\beta$-decay as a case of Yb.
The N(s)/N(p) ratios for Dy calculated by the various stellar models
are systematically larger than those of other $p$-nuclei (see Fig.~\ref{fig:sp_cal}).
This can be understood by effects of the $\beta$-decay from the second path.
The $s$-nucleus $^{158}$Dy is strongly populated by the $\beta$-decay 
but the $p$-nucleus $^{156}$Dy is dominantly produced by the first path (see Fig.~\ref{fig:paths} (b)).
As the result, the N($^{158}$Dy)/N($^{156}$Dy) ratio increases after the $\beta$-decay.
In contrast, the calculated ratios of Er are systematically lower than the average value.
The $p$-nucleus $^{164}$Er is populated by the $\beta$-decay
but the $s$-nucleus $^{166}$Er is not populated
(see Fig.~\ref{fig:paths} (b)).
As the result, the N($^{166}$Er)/N($^{164}$Er) ratio decreases after the $\beta$-decay.
The populations by the $\beta$-decay are determined by the position and width of the second path,
which depend strongly on the ($\gamma$,$\alpha$) reaction rates.
Thus, the final N(s)/N(p) ratios 
are sensitive to the ($\gamma$,$\alpha$) reaction rates on neutron-deficient isotopes.

Rauscher (2006) calculated photodisintegration reaction rates with various nuclear models
and presented branching points where
the ($\gamma$,$\alpha$) or ($\gamma$,p) reaction rate competes against the ($\gamma$,n) reaction rate.
The results showed that the branching points
depend on the nuclear models. 
The discussion mentioned above shows that 
the actual ($\gamma$,$\alpha$) reaction rates on Hf and Yb isotopes
may be lower than the reaction rates used in the present calculations
and the second path may be located far from the $\beta$-stability line.
If so, both the p- and s-nuclei of Dy and Er are populated by the $\beta$-decay as the case of Yb
and the final N(s)/N(p) ratios are not largely changed from those before the $\beta$-decay.
Thus, the reaction rates based on experimental data with high accuracy
are desired.
The ($\gamma$,$\alpha$) reaction rates have been systematically studied 
using $\alpha$ particle induced reactions \citep{alpha06,Kiss06,alpha07}.
A new generation of $\gamma$-ray sources
of a laser Compton scattering (LCS) $\gamma$-rays in an energy range of MeV has been developed \citep{Duke,AIST,NewSUBARU}
and widely used for studies of nuclear astrophysics \citep{Utsunomiya03,Mohr04,Shizuma05,Hayakawa06b}.
The LCS sources are powerful tools for measuring nuclear properties as photo-induced reaction rates
because of the sharp energy edge in their energy spectra and the tunable energy.
The measurements of ($\gamma$,$\alpha$) reaction rates on the $p$-nuclei using the LCS $\gamma$-rays
will contribute to the improvement of $\gamma$-process models \citep{Utsunomiya06}.

\subsection{\it Extended universality}

The first scaling leads to the constant N(s)/N(p) ratios against the atomic number
in individual $\gamma$-processes
but their absolute values are considered to depend strongly on astrophysical conditions.
However, most N(s)/N(p) ratios calculated in the four different SN models
are not only constant but also centered on a specific value of 3 within a factor of 3 (see Figs.~\ref{fig:sp_cal}).
These results suggest an extended universality that the N(s)/N(p) ratios produced by individual SN 
$\gamma$-processes are centered on the specific value of 3.

\subsection{\it Frequency of the $s$-process in AGB stars and massive stars in the Galaxy}

\begin{figure}
\includegraphics[viewport=0mm 0mm 210mm 120mm, clip, scale=0.5]{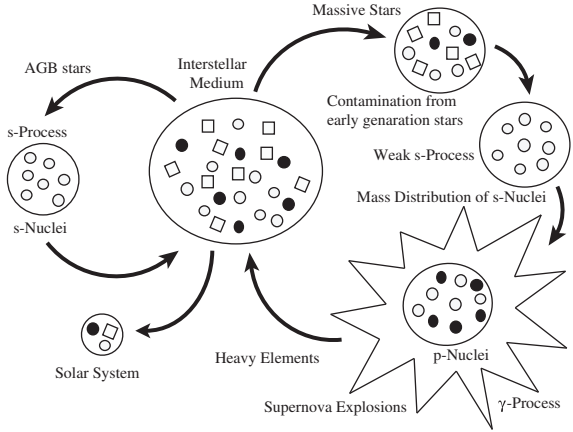}
\caption{The schematic chart for the galactic chemical evolution for the p-, s- and s-nuclei.
The s-nuclei in the ISM are dominantly produced in the main s-process in the AGB stars.
The massive stars have contamination of heavy elements from the ISM.
The heavy elements are irradiated by neutrons in weak s-process before SN explosions
and the abundance distribution of the heavy elements is changed to that of the s-process.
The $p$-nuclei are synthesized from the $s$-processed seed
by the $\gamma$-process at the supernova explosion. 
The $s$-process depends strongly on
the metallicity. The N(s)/N(p) ratio is proportional to the frequency of the s-process
events and is time dependent.}
\label{fig:ISM}       
\end{figure}

The universality of the first scaling is important for a study of
a frequency of the $s$-process episodes in the AGB stars and massive stars.
Figure \ref{fig:ISM} shows a schematic chart for the Galactic chemical evolution (GCE).
The $s$-nuclei in the solar system originate mainly from the main $s$-process in the AGB stars \citep{Gallino98,Busso99}.
The abundance, N$_{\odot}$(s), is proportional to the abundance synthesized by the individual main $s$-processes
and the frequency of the formation of the AGB stars,
whereas N$_{\odot}$(p) is proportional to the $\gamma$-process abundance  and the frequency of SNe.
Using  the estimated frequency of SNe and the calculated abundances
of the $p$- and $s$-nuclei in single nucleosyntheses,
we can estimate the frequency of the AGB $s$-process relative to the $\gamma$-process.

The extended universality gives a ratio of $s$-nucleus abundance
originated from the AGB stars to that from the massive stars.
The extended universality 
shows that the N$_{SN}$(s)/N$_{SN}$(p) ratios produced by a single SN
is almost equal to 3 in the mass region of A$>$90.
The $s$-process abundances in the solar system are contributed from both the AGB stars and massive stars
associated with SNe;
namely, N$_{\odot}$(s) = N$_{AGB}$(s)+N$_{SN}$(s).
The $p$-nuclei originate dominantly from SNe; namely, N$_{SN}$(p) = N$_{\odot}$(p).
Using the first scaling of N$_{\odot}$(s)/N$_{\odot}$(p) = 23
and the extended universality of N$_{SN}$(s)/N$_{SN}$(p) = 3,
we obtain a result that the ratio of N$_{AGB}$(s)/N$_{SN}$(s) at the solar system is about 6.7
in the mass region of A $>$ 90.

\subsection{\it Proposal for astronomical observations of Indium to verify the extended universality}

The verification  of the extended universality of the $\gamma$-process
is essential for the discussion mentioned above.
Recent progress in spectroscopic studies of metal-poor stars has
enabled isotope separation
of several heavy elements such as Eu 
by measurements of hyperfine splitting and isotope shifts \citep{Sneden02,Lambert02,Aoki03b}.
The observation of $^{113}$In 
is most promising among all the $p$-nuclei in extra-solar objects
because this element has the following two advantages.
The first is that this element has only two odd-$N$ stable isotopes,
a $^{113}$In $p$-nucleus, and a $^{115}$In $s$-nucleus.
The second is that the $^{113}$In $p$-nucleus has an isotopic fraction as large as 4.29~\%
in the solar system.
In contrast, most $p$-isotopes are even-even nuclei and have many isotopes.
There are three other odd-$Z$ or odd-$N$ $p$-nuclei; $^{115}$Sn, $^{138}$La and $^{180}$Ta,
but their isotopic fractions are small (0.012 $\sim$ 0.34 \%).
The absorption line of In (4511.31$\AA$) was detected in the solar photosphere \citep{Goldberg60,Lambert69}.
Recently, the In $_{I}$ line at 4511 $\AA$ in sun-like stars were detected \citep{Gonzalez06}.
If the indium isotopic fractions could be observed in metal-deficient r-process enhanced stars,
it would be valuable evidence for the universality of the $\gamma$-process.
The calculated N($^{115}$In)/N($^{113}$In) ratios in the four models are 3.1 $\sim$ 8.5
corresponding to the $^{113}$In fraction of 11\% $\sim$ 24\%.
The fraction of the $p$-isotope in material affected strongly by a single SN
should be enhanced,
although the enhancement may be small because of the mixing of materials from other layers.

\section{A piece of evidence for the weak $s$-process}

We have presented that the weak $s$-process  before SN explosions
is important for the proportionality between the $p$- and $s$-nuclei
in the solar abundances.
The three large deviations of  N$_{\odot}$(s)/N$_{\odot}$(p) for Ce, Er, W
give a hint to understand the weak $s$-process.
The ratios of the three elements are
larger by an order of magnitude than those of the other elements (see Fig.~\ref{fig:sp_solar}).
These deviation originate from enhancements of heavier isotopes,$^{140}$Ce, $^{166}$Er and $^{182}$W,
in their pairs.
These three isotopes are not shielded against the $\beta$ decay after the freezeout of the $r$-process
and the contributions of the $r$-process for $^{166}$Er and $^{182}$W
are larger than those of the $s$-process \citep{Arlandini99}.

The solar abundances can be expressed by
\[
N_{\odot}(Z,N)=N_{p}(Z,N)+N_{s}(Z,N)+N_{r}(Z,N).
\]
The abundances of typical $s$-nuclei are expressed by N$_{\odot}$(Z,N) = N$_{s}$(Z,N),
but the abundances of the three isotopes
are expressed by N$_{\odot}$(Z,N) = N$_{s}$(Z,N)+N$_{r}$(Z,N).
The first scaling indicates that the abundance of the $p$-nucleus, N$_{p}$(Z,N-2) or N$_{p}$(Z,N-4),
is proportional to the heavier isotope abundance of N$_{s}$(Z,N)+N$_{r}$(Z,N) at SN explosions;
 namely N$_{p}$(Z,N-2) = $\alpha$$\cdot$(N$_{s}$(Z,N)+N$_{r}$(Z,N)),
where $\alpha$ is about 1/23 at the solar system.
The abundances of the $p$-nuclei can be expressed by
 N$_{p}$(Z,N-2) =  N$_{p}^{r}$(Z,N-2) + N$_{p}^{s}$(Z,N-2),
where N$_{p}^{r}$(Z,N-2) = $\alpha$$\cdot$N$_{r}$(Z,N),
and N$_{p}^{s}$(Z,N-2) = $\alpha$$\cdot$N$_{s}$(Z,N).

The weak $s$-process occurs in the massive stars before the SNe
and the mass distribution of pre-existing heavy elements
is changed to that of the $s$-process (see Fig.~\ref{fig:ss}).
The abundance N$_{s}$(Z,N)+N$_{r}$(Z,N) is changed to N$_{s}$(Z,N)
and the $p$-nucleus abundance N$_{p}^{r}$(Z,N-2) originated from N$_{r}$(Z,N) vanishes;
for example, N$_{p}$($^{164}$Er) = 1/23$\cdot$N$_{s}$($^{166}$Er).
Because the solar abundances of $^{140}$Ce, $^{160}$Er, and $^{182}$W are
contributed from both the $r$- and $s$-processes,
the N$_{\odot}$(s)/N$_{\odot}$(p) ratios in the three elements
are expressed by (N$_{s}$(Z,N)+N$_{r}$(Z,N))/N$_{p}$(Z,N-2),
which are larger than N$_{s}$(Z,N)/N$_{p}$(Z,N-2) of neighboring $p$-nuclei
by a factor of (N$_{s}$(Z,N)+N$_{r}$(Z,N))/N$_{s}$(Z,N).
In this way, the three large enhancements in N(s)/N(p) ratios 
are observed in the solar abundances.
In contrast, if the weak $s$-process does not occur before the $\gamma$-process,
the abundances of the $p$-nuclei in the three elements
are also proportional to those of heavier isotopes.
Therefore, 
the three deviations for Ce, Er and W are a piece of evidence that
the weak $s$-process in the massive stars occurs before the $\gamma$-process.

\subsection{\it Thermometer for the $s$-process}

\begin{figure}
\includegraphics[viewport=0mm 0mm 140mm 70mm, clip, scale=0.7]{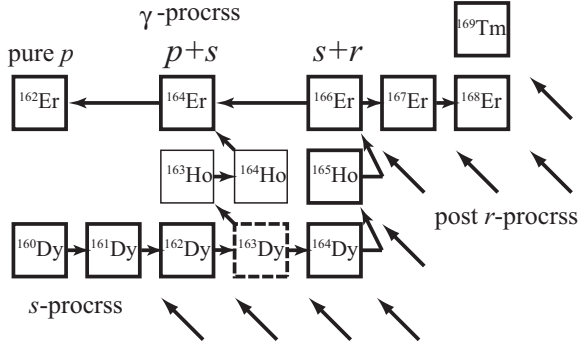}
\caption{A partial nuclear chart around Er and Dy isotopes and nucleosynthesis flows.
The second $p$-nucleus $^{162}$Er is dominantly synthesized by the $\gamma$-process.
In contrast, the first $p$-nucleus $^{164}$ is synthesized by the $\gamma$- and $s$-processes.
In high temperature environments of T $\sim$ 10$^8$ K, $^{163}$Dy becomes unstable
by atomic effects and decays to $^{163}$Ho. $^{166}$Er is synthesized by the $s$- and $r$-processes
since it is not shielded against the $\beta$-decay after the $r$-process.}
\label{fig:Er_Ho}       
\end{figure}

A question why the abundance of $^{164}$Er is an order of magnitude larger
than those of neighboring $p$-nuclei as shown in Fig.~\ref{fig:abundances} 
has been known about fifty years ago \citep{BBFH}.
Takahashi and Yokoi (1983) suggested that a stable isotope $^{163}$Dy becomes to unstable to 
${\beta}^-$-decay in a typical $s$-process temperature
because of an atomic effect of ionized ions
and $^{164}$Er may be synthesized by a weak branch of the $s$-process,
\[
\mbox{
$^{163}$Dy$^*(e^- \nu)^{163}$Ho(n,$\gamma$)$^{164}$Ho$(e^-\nu)^{164}$Er.
}
\]
The $\beta$-decay rate of $^{163}$Dy depends on thermodynamic condition of the $s$-process,
and hence the abundance ratio of $N_{s}$($^{166}$Er)/$N_{s}$($^{164}$Er) is sensitive to the temperature.
However, the evaluation of $N_{s}$($^{166}$Er)/$N_{s}$($^{164}$Er) seems to be difficult,
because $^{164}$Er is produced by the $s$- and $\gamma$-processes
and $^{166}$Er is produced by the $s$- and $r$-processes as shown in Fig.~\ref{fig:Er_Ho}.
We here propose a new method to evaluate the $s$-process abundances of $^{164,166}$Er
using the empirical scalings; N(s)/N(p) = 23 and N(1st p)/N(2nd p) = 1.
Because the second $p$-nucleus $^{162}$Er is dominantly synthesized by the $\gamma$-process,
we can evaluate the $p$-process abundance of $^{164}$Er and the seed $s$-nucleus abundance of $^{166}$Er
using the scalings.
The evaluated ratio, N$_s$($^{166}$Er)/ N$_s$($^{164}$Er), is 2.2 at the solar system.
Arlandini {\it et al}.~(1999) calculated the component of all $s$-nuclei
using a stellar model for low-mass AGB stars with $^{13}$C burning.
The calculated ratio N$_s$($^{166}$Er)/ N$_s$($^{164}$Er) is 3.8,
which is consistent with the present evaluated ratio of 2.2
within a factor of 2.
Therefore, the empirical scalings support the $s$-process calculations by Arlandini et al.

\section{OTHER ORIGINGES}

Our discussion shows that
the two scalings are a piece of evidence that twenty-seven $p$-nuclei
are dominantly produced by the SN $\gamma$-process.
Another model discussed in the literature \citep{Howard91}
is $\gamma$-process in the outer layers of exploding white dwarf stars. 
Those outer layers should have been enriched in s-process nuclei as the the ashes 
of He shell burning settled on them in the star's AGB phase.  
If the burning front passing through them during the explosion tends 
to give the right set of conditions, a $\gamma$-process might occur.
Though the p-isotope yields from exploding white dwarf stars are highly uncertain, 
they can be enormous and may be significant contributors to the solar system's 
supply of the p-isotopes. 
The question whether the $\gamma$-process in the white dwarf stars
can satisfy the scaling laws has been remained.

There are thirty-five $p$-nuclei and the origin of the other eight $p$-nuclei
has remained as an open question.
The other $p$-nuclei are $^{92,94}$Mo, $^{96,98}$Ru, $^{144}$Sm, $^{115}$Sn, $^{138}$La, and $^{180}$Ta.

\subsection{Even-even nuclei}

The first scaling at the solar abundances shows large deviation for Mo and Ru isotopes (see Fig.~\ref{fig:sp_solar}).
The isotope abundances of these nuclei are about ten times larger than those of typical $p$-nuclei.
It has been well known that the $\gamma$-process calculations underproduce
the abundances of these isotopes, $^{92,94}$Mo and $^{96,98}$Ru.
Recently, as the origin of these isotopes,
nucleosynthesis in early neutrino wind in core-collapse SNe ($\nu$$p$-process)
have been studied \citep{Pruet06,Frohlich06,Wanajo06}.

Burbidge et al. (1957) pointed out that the mass distribution of the $p$-nuclei
show two peaks near two neutron magic numbers of N=50, 82 and
the $p$-nuclei may be synthesized by (p,$\gamma$) and ($\gamma$,n) reactions.
The abundance pattern around Mo region
is similar to that of Sm.
As shown in Fig.~\ref{fig:abundances}, the abundances of $^{92,94}$Mo
is larger than those of lighter $p$-isotopes, $^{84}$Sr and $^{78}$Kr,
and the abundance of $^{144}$Sm is also larger than those
of lighter $p$-nuclei.
In contrast, the abundances of the seed $s$-nuclei, $^{96}$Mo, $^{100}$Ru
and $^{148}$Sm, are consistent with those of the same mass region:
namely, their abundances are lower than those of $s$-nuclei in the lighter mass regions
beyond the neutron magic numbers.
This fact suggests a possibility that these five $p$-nuclei, $^{92,94}$Mo, $^{96,98}$Ru and $^{144}$Sm
might be synthesized mainly from $s$-nuclei located in lighter mass regions
beyond the neutron magic number
by particle-induced reactions such as (p,$\gamma$),($\alpha$,$\gamma$) or
(n,$\gamma$) reaction.
Recently, the proton capture experiments have been carried out
up to $A$ $\sim$ 120 \citep{Ozkan02,Spyrou07}.
However, the reaction rates in a heavy mass region of $A$ $\sim$ 140
have not been measured. These rates are of importance
for understanding the origin of the p-nuclei near both the neutron
magic numbers of $N$ = 50 and 82.

Recent progress of meteorite science give crucial hints about the origin of Mo and Sm.
Yin et al. (2002) measured isotopic fractions of Mo in primitive meteorites with high accuracy
and isotopic fractions of the $p$-nucleus $^{94}$Mo and the $r$-nucleus $^{100}$Mo show
different anomalies in comparison with the solar abundances.
Yin et al. concluded that there are at least three patterns of the isotopic anomalies of the Mo isotopes.
This result suggests that the nucleosynthesis site of $^{94}$Mo
is not correlated with the $r$-process site.
Isotopic anomalies of Sm in primitive meteorites were also measured
and the results showed that nucleosynthesis sites supplying the p- and r-isotopes were disconnected
or only weakly connected \citep{Sm}.
It is of importance measurements of the isotope abundances of both the elements, Sm and Mo, 
in individual meteoritic samples,
which can answer the question whether the nucleosynthesis sites of Sm and Mo are same or not.

\subsection{Odd-$A$ nuclei}

As the origin of $^{115}$Sn, the $\gamma$-,
$s$- and $r$-processes were proposed.
$^{115}$Sn may be synthesized through a $\beta$ unstable isomer in $^{115}$In
after the freezeout of the $r$-process
and by a weak branch of the $s$-process
by
\[
\mbox{
$^{113m}$Cd$(e^- \nu)^{113}$In(n,$\gamma$)$^{114}$In$(e^-\nu)^{114}$Sn(n,$\gamma$)$^{115}$In.
}
\]
However, previous studies reported that $^{115}$Sn cannot be produced enough
by the $s$-process \citep{Nemeth94}.
M{\'e}meth et al. suggested that 
the population of the isomer in $^{113}$Cd
should follow by thermal-equilibrium conditions
in the s-process environment $kT$=12-30 keV.
In such the case, the population ratio of the isomer depend only on
the temperature of the environment.
Here we would like to point out that
recent studies of the $s$-process suggest a possibility that the s-process may occur
at the low temperature $kT$ $<$ 10 keV.
In such the case, 
the population of the excited states in $^{113}$Cd dose not follow by the thermal-equilibrium conditions
and thus the $^{115}$Sn abundance depends on a partial neutron capture
reaction cross-section to the isomer of $^{113}$Cd.
The neutron capture reaction cross-section to the $^{113}$Cd isomer
has never measured with high accuracy at any energy
and the measurement is desired.

\subsection{Odd-odd nuclei}

Two isotopes, $^{138}$La and $^{180}$Ta, 
are classified to a group of odd-odd nuclei.
These two isotopes have the following unique features.
First their isotope abundances
are small, 0.090\% (for $^{138}$La) and 0.012\% ($^{180}$Ta).
Second they are shielded against both of ${\beta}^+$-decay and ${\beta}^-$-decay
by stable isobars and thus they are produced only by direct nuclear reactions,
for example ($\gamma$,n) reactions from heavy isotopes or
(${\nu}_e$,e$^-$) reactions from isobars.

$^{138}$La: As the origin of $^{138}$La, the cosmic-ray process \citep{Audouze70}
and the $\nu$-process \citep{Woosley90} were proposed.
A result that about 90\% in the solar abundance of $^{138}$La
can be explained by the $\nu$ process was reported \citep{Heger05}
and the contribution of a key charged current reaction to $^{138}$La
was quantitatively evaluated \citep{Byelikov07}.
A study of primitive meteorites provides a hint about the origin of $^{138}$La.
Shen and Lee (2003) reported the isotope abundance anomalies
of several elements as Ti and La in Ca-Al-rich inclusions,
which is considered to originated from the early solar nebula.
The abundance ratios of $^{138}$La/$^{139}$La
is correlated with that of $^{50}$Ti/$^{48}$Ti \citep{Shen03}.
The reason why $^{138}$La is correlated with $^{50}$Ti has been an open question.

$^{180}$Ta: The origin of $^{180}$Ta is one of hot topics
in the nuclear astrophysics
since $^{180}$Ta has a unique feature that 
the meta-stable state in $^{180}$Ta is the only isomer (half-life 
of $\ge$ 10$^{15}$ yr) existing in the solar system, 
while the ground state is unstable state against $\beta$ -decay (half-life of 8 hr).
As the origin of $^{180}$Ta,
the $s$-process \citep{Yokoi83,Belic99},
and the $\gamma$-process \citep{Woosley78},
the $\nu$-process \citep{Woosley90} and the cosmic-ray process \citep{Audouze70}
were suggested.
A recent $\nu$-process calculation overproduces the solar abundance of $^{180}$Ta
relative to $^{16}$O, which based on a new experimental result using
a $^{180}$Hf($^{3}$He,t)$^{180}$Ta reaction 
to evaluate the charged current reaction rate \citep{Byelikov07}.

In explosive nucleosynthesis such as the $\gamma$- and $\nu$-processes,
the population ratio of the meta-stable isomer to the $\beta$-unstable ground state
at freezeout of a nucleosynthesis process
is an important factor.
The transition probability from the isomer to the ground state
was measured using real photons \citep{Belic99}
and the population ratio by a ($\gamma$,n) reaction
was also measured \citep{Goko06}.
To estimate exactly the residual abundance of $^{180}$Ta,
a time-dependent calculation is required.
However, for the estimation of the isomer population,
a thermal equilibrium approximation \citep{Rauscher02} are widely used, wherein
the population ratio of the isomer to the ground state can be calculated
by the Maxwell-Boltzmann population at critical temperature $T_{crit}$
that is the lowest temperature of the thermal equilibrium region.

It should be noted that
exited states located above the isomer
are also populated in high temperature environments
and may decay to the isomer.
On other words, the excited states above the isomer 
also contribute the residual abundance of $^{180}$Ta.
Mohr et al. (2007) calculated the temperature of $T_{crit}$ based on measured transition
probability \citep{Belic99}
and concluded $T_{crit}$ is 40.4 keV which leads to $P_{m}/P_{total}$ of 0.35.

\begin{figure}
\includegraphics[viewport=0mm 0mm 140mm 110mm, clip, scale=0.75]{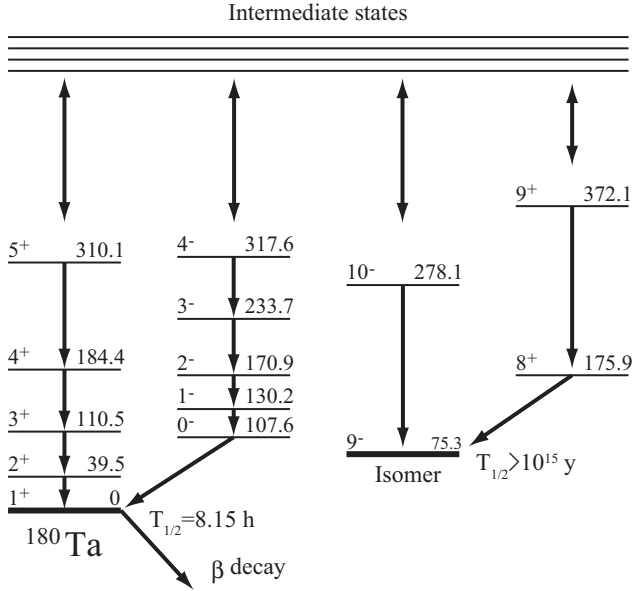}
\caption{A partial nuclear level scheme of $^{180}$Ta. 
$^{180}$Ta is excited in a high temperature environment of $T$ $>$ 10$^8$ K.
The ground state and excited states are populated from with each other 
via intermediate states lying higher excited states.
After a supernova explosion, the temperature of a nucleosynthesis site
suddenly decreases. The excited states in $^{180}$Ta
decay to the ground state or the isomer.
The excitation energies
are taken from Ref.~\citep{Saitoh99}.}
\label{fig:180Ta}       
\end{figure}

Here we discussed the population of the isomer under the thermal equilibrium approximation.
A population of an excited state is followed to the Maxwell-Boltzmann population presented by
\begin{equation}
P_{i}/P_{gs} = (2J_{i}+1)/(2J_{gs}+1)exp(-E_{i}/kT).
\label{eq:population}
\end{equation}
In a temperature range of $T^{9}$ = 0.1-1.0,
all excited states lower than a few hundred keV energy
are populated (see Fig.~\ref{fig:180Ta}).
After the freezeout, the each excited state decays to
the ground state or the isomer.
The final population ratio of the isomer to the total
is presented by
\begin{equation}
P_{m}/P_{total} = \frac{ {\sum}(P_{i}/P_{gs})}
{ {\sum}(P_{i}/P_{gs})+{\sum}(P_{j}/P_{gs}) }
\label{eq:isomer_ratio}
\end{equation}
where $i$ and $j$ indicate levels decaying to the isomer and
the ground state, respectively.
We calculate the contribution of the excited levels presented in Fig.~\ref{fig:180Ta},
since the contribution of the higher levels is lower than 0.1\%.
In previous calculations, the residual ratio of 0.3-0.5 have been used \citep{Rauscher02}.
However, the present calculated ratios are 0.06, 0.13, 0.20, and 0.26 
for the assumed critical temperature of $T_{crit}^{9}$ = 0.20, 0.25, 0.30, and 0.35, respectively.
This result shows that the residual abundance of $^{180}$Ta
should decrease against the previous estimations
if the thermal equilibrium is a good approximation.

\section{CONCLUSION}

We present two empirical scaling laws concerning the
$p$- and $s$-nuclei with the same atomic number in the solar system abundances.
The first $p$-nucleus and the second $p$-nucleus are lighter than the $s$-nucleus
by two and four neutrons, respectively.
The first scaling is the correlation between the $p$- and $s$-nuclei.
The corrected N$_{\odot}(s)$/N$_{\odot}$(1st p) and N$_{\odot}(s)$/N$_{\odot}$(2nd p)
ratios are almost constant over a wide region of the atomic number
and are mostly centered around 23 within a factor of 2.
These scalings are a piece of evidence that the $\gamma$-process
is the most probable origin of the twenty-seven $p$-nuclei.

The scalings in the solar system lead
to a novel concept "the universality of the $\gamma$-process"
that the abundance ratios N(s)/N(p) produced
by individual SN $\gamma$-processes are almost constant over
a wide region of the atomic number.
The $\gamma$-process calculations of core-collapse SNe under various astrophysical conditions
support the universality of the $\gamma$-process.
Three mechanisms contribute to the universality.
The first mechanism is the weak $s$-process. 
It is well known that light elements of $A$ $<$ 90 are produced by successive neutron capture
reactions from iron seeds in the weak $s$-process. 
However, pre-exiting heavy elements of $A$ $>$ 90 are also irradiated
by neutrons and their mass distribution is changed locally to that of the main $s$-process
in asymptotic giant branch stars.
Therefore, the mass distribution patterns of the seed nuclei at the supernova explosions are almost identical.
The second mechanism is the shifts of the mass regions of the $\gamma$-process layers to keep their peak 
temperature. 
The $\gamma$-process layers are shifted to outer layers 
as the explosion energy or mass increases. Inside the $\gamma$-process layers, the seed nuclei
are completely destroyed by photodisintegration reactions, 
whereas the $p$-nuclei outside the $\gamma$-process layers 
are not synthesized because of their low temperature.
Therefore, the physical conditions of the $\gamma$-process layers are almost identical.
The third mechanism is independence from two nucleosynthesis flows.
The $p$-nuclei are produced by the two flows:
direct ($\gamma$,n) reactions from heavier isotopes, 
and $\beta$-decay after downflow by ($\gamma$,p)
and ($\gamma$,$\alpha$) reactions from heavier elements
which may break the scalings.
We calculated the contributions of both the flows for each $p$-nucleus.
The ($\gamma$,n) reactions are dominant in the light mass region of $N$ $<$ 82.
The $\beta$-decay after the downflow contributes to the $p$-nucleus abundances 
in the heavy mass region of $N$ $>$ 82
but does not drastically change the abundance ratios
because both the p- and s-nuclei are populated by $\beta$-decay in most cases.
In this way, the scalings hold for individual $\gamma$-processes.

In addition, we find that large deviations of N$_{\odot}$(s)/N$_{\odot}$(p)
for Ce, Er, and W in the solar system
are a piece of evidence that the weak $s$-process actually occurred before SNe.
The present calculated results suggest an extended universality that the N(s)/N(p) ratios 
produced in individual SNe may be constant around a specific value of 3
independent of the stellar conditions.
With the extended universality of N$_{SN}$(s)/N$_{SN}$(p) = 3 and the first scaling
of N$_{\odot}$(s)/N$_{\odot}$(p) = 23, we estimate that the ratio of the $s$-process abundance contribution from
the AGB stars to the massive stars
associated with SNe is about 6.7 for the $s$-nuclei of $A$ $>$ 90 in the solar system.
An astronomical observation with indium isotope separation
is useful for the verification of the extended universality.
This observation can give a crucial impact on the identification of the $\gamma$-process site
and GCE for $p$- and $s$-nuclei.

\noindent
{\bf Table.1} List of the p-process isotopes. The isotope abundances are taken from Bi{\'e}vre and Taylor (1993).
(a) the contribution of the r-process. (b) the contribution of the s-process. (c) the effect of the neutron magic number.\\
\begin{tabular}{llllllllllll}
\hline
    &Z  &	2nd p  &1st p  &s  &N(2nd p)  &	N(1st p)  &	N(s)  &	N(s)      &N(s)     &N(2nd p)  & \\
         &   &         &       &   &          &           &           & /N(2nd p) &/N(1st p) &/N(1st p) & \\
\hline
Se  &	34  &  &	74  &	76  &  &	0.89  &	9.36  &  &	10.5 	  &  & \\
Kr  &	36  &  &	78  &	80  &  &	0.35  &	2.25  &  &	6.43 	  &  & \\
Sr  &	38  &  &	84  &	86  &  &	0.56  &	9.86  &  &	17.6 	  &  & \\
Mo  &	42  &	92  &	94  &	96  &	14.84&  9.25  &	16.68  &1.12 &	1.80  &	0.62   & \\
Ru  &	44  &	96  &	98  &	100  &	5.52&	1.88  &	12.6  &	2.28 &	6.70  & 0.34   & \\
Pd  &	46  &  &	102  &	104  &  &	1.02  &	11.14  &  &	10.9 	  &  & \\
Cd  &	48  &	106  &	108  &	110  &	1.25&	0.89  &	12.49  &9.99 &	14.0  & 0.71  & \\
In  &	49  &  &	113  &	115  &  &	4.3  &	95.7$^{(a)}$  &  &	22.3$^{(a)}$   &  & \\
Sn  &	50  &	112  &	114  &	116  &	0.97&	0.65  &	14.53  &15.0 &	22.4  & 0.67   & \\
Sn  &	50  &  &	115  &	     &	&	0.34  &  &  &			  &  & \\
Te  &	52  &  &	120  &	122  &  &	0.096  &2.603  &  &	27.11 	  &  & \\ 
Xe  &	54  &	124  &	126  &	128  &	0.10&	0.09  &	1.91  &	19.1 &	21.2 	  &0.90   & \\
Ba  &	56  &	130  &	132  &	134  &	0.106&	0.101  &2.417  &22.80 &	23.93 	  &0.95   & \\
La  &	57  &  &	138  &       &	&	0.0902  &  &  &			  &  & \\ 
Ce  &	58  &	136  &	138  &	140  &	0.19&	0.25  &	88.48$^{(a,c)}$  &466$^{(a,c)}$ &354$^{(a,c)}$  & 1.32   & \\
Sm  &	62  &	144  &	     &	148  &	3.1&	  &   11.3 &  &	3.65 		  &  & \\
Gd  &	64  &  &	152  &	154  &  &	0.20$^{(b)}$  &	2.18  &  &	10.9$^{(b)}$ 	  &  & \\
Dy  &	66  &	156  &	158  &	160  &	0.06&	0.10  &	2.34  &	39.0 &	23.4 	  &1.67   & \\
Er  &	68  &	162  &	164  &	166  &	0.14&	1.61$^{(b)}$  &	33.6$^{(a)}$  &	240$^{(a)}$ &	20.9$^{(a,b)}$ 	  &11.5   & \\
Yb  &	70  &  &	168  &	170  &  &	0.13  &	3.05  &  &	23.5 	  &  & \\
Hf  &	72  &  &	174  &	176  &  &	0.162  &5.206  &  &	32.14 	  &  & \\
Ta  &	73  &  &	180  &       &	&	0.012  &  &  &	  	  &  & \\
W  &	74  &  &	180  &	182  &  &	0.13  &	26.3$^{(a)}$  &  &	202$^{(a)}$ 	  &  & \\
Os  &	76  &  &	184  &	186  &  &	0.02  &	1.58  &  &	79 	  &  & \\
Pt  &	78  &  &	190  &	192  &  &	0.01  &	0.79  &  &	79 	  &  & \\
Hg  &	80  &  &	196  &	198  &  &	0.15  &	9.97  &  &	66.5 	  &  & \\

\hline
\end{tabular}

\end{document}